\documentclass[12pt]{iopart}
\usepackage{iopams,graphicx}
\begin{document}

\title[Pair Correlations and Susceptibilities in Exactly Solvable Models]
{Some Recent Results on Pair Correlation Functions
and Susceptibilities in Exactly Solvable Models}

\author{Jacques H H Perk and Helen Au-Yang}

\address{Department of Physics, Oklahoma State University, 
145 Physical Sciences, Stillwater, OK 74078-3072, USA}
\ead{perk@okstate.edu}
\begin{abstract}
Using detailed exact results on pair-correlation functions of
$Z$-invariant Ising models, we can write and run algorithms of polynomial
complexity to obtain wavevector-dependent susceptibilities for a variety
of Ising systems. Reviewing recent work we compare various periodic and
quasiperiodic models, where the couplings and/or the lattice may be
aperiodic, and where the Ising couplings may be either ferromagnetic, or
antiferromagnetic, or of mixed sign. We present some of our results on the
square-lattice fully-frustrated Ising model. Finally, we make a few
remarks on our recent works on the pentagrid Ising model and on
overlapping unit cells in three dimensions and how these works can be
utilized once more detailed results for pair correlations in, {\it e.g.},
the eight-vertex model or the chiral Potts model or even three-dimensional
Yang--Baxter integrable models become available.
\end{abstract}

\vspace{2pc}

\section{Introduction}
Detailed exact results on pair-correlation functions and susceptibilities
are usually hard to come by, even in models that are ``exactly solvable."
The best results are available for $Z$-invariant Ising-type models, and
even though the results are often not in closed form, we may follow Tony
Guttmann and claim an exact result if we can produce an algorithm of
polynomial complexity to obtain detailed information on, say,
wavevector-dependent susceptibilities for a variety of Ising systems.

For a long time, the paper of Wu, McCoy, Tracy and Barouch \cite{WMTB76}
has been the standard work on the pair correlations and susceptibility
of the square-lattice Ising model. Only in the last few years do we know
how to do much better. We now have algorithms to construct long high-
and low-temperature series together with detailed expansions at
criticality for the susceptibility \cite{ONGP1,ONGP2} using quadratic
recurrence relations \cite{P1980} for pair correlations. We also have
much more analytic and numerical information for pair correlations
and susceptibilities in more general {\it Z}-invariant inhomogeneous
Ising models \cite{AJP,APMO1,APMO2}, together with new understanding from
a field theory approach \cite{CHPV}.

\subsection{Baxter's {\it Z}-invariant inhomogeneous Ising model}

We can start with an inhomogeneous Ising model on a square lattice
with reduced interaction energy
\begin{equation}
-\beta{\cal H}\,=\sum_{m,n}\,({\bar K}_{m,n}\sigma_{m,n}\sigma_{m,n+1}
+ K_{m,n}\sigma_{m,n}\sigma_{m+1,n})
\end{equation}
where $\beta=1/k_{\rm B}T$ the inverse temperature. At site $(m,n)$ the
spin takes values $\sigma_{m,n}=\pm1$, while $K=\beta J$ and
${\bar K}=\beta{\bar J}$ are ``horizontal" and ``vertical"
dimensionless coupling constants.

One important quantity to study is the wavevector-dependent susceptibility
$\chi({\bf q})$ defined by
\begin{equation}
k_{\rm B}T\chi({\bf q})\equiv\bar\chi({\bf q})=
\lim_{{\cal L}\to\infty}{\frac 1 {{\cal L}}}
\sum_{\bf r}\sum_{\bf r'} {\rm e}^{{\rm i}{\bf q}\cdot({\bf r'}-{\bf r})}
\big[{\langle\sigma_{\bf r}\sigma_{\bf r'}\rangle}-
\langle\sigma_{\bf r}\rangle\langle\sigma_{\bf r'}\rangle\big]
\label{chiq}
\end{equation}
This is the Fourier transform of the connected pair correlation function.
In (\ref{chiq}), ${\cal L}$ is the number of lattice sites, ${\bf r}$ and
${\bf r}'$ run through all sites and ${\bf q}=(q_x,q_y)$ is the
wavevector.

\begin{figure}[tbh]
\begin{center}
\includegraphics[width=0.6\hsize]{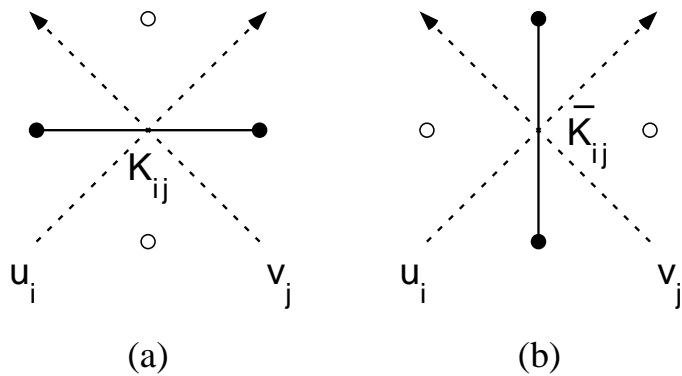}
\end{center}

\caption{Two oriented rapidity lines with rapidities $u_i$ and $v_j$ are
drawn dashed. The orientation of the coupling with respect to these
rapidity lines defines whether we have (a) horizontal coupling
$K_{ij}=K(u_i,v_j)$ or (b) vertical coupling
$\bar K_{ij}=\bar K(u_i,v_j)$.}
\label{fig1}
\end{figure}

Baxter's {\it Z}-invariant Ising model \cite{BaxZI,BaxFF,AP-ZI} is defined
in terms of oriented rapidity lines on which the rapidity variables live.
(The rapidity lines can be moved at will, without changing the partition
function $Z$, so we do not have to take a square lattice.) The faces
bounded by the rapidity lines are alternatingly colored black or white and
spins live somewhere on the black faces forming the Ising system
$\{\sigma_{\bf r}\}$, whereas dual spins live on the white faces forming
the dual Ising system $\{\sigma^{\ast}_{\bf r^{\ast}}\}$. Spins interact
only with nearest neighbours from which they are separated by precisely two
rapidity lines, see Fig.~\ref{fig1}.
The couplings $K$ and $\bar K$ are parameterized in terms of elliptic
functions of modulus $k$, {\it i.e.},
\begin{eqnarray}
\sinh\big(2K(u_1,u_2)\big)&=k\,{\rm sc}(u_1-u_2,k')
={\rm cs}\big({\rm K}(k')+u_2-u_1,k'\big)\cr\cr
\sinh\big(2{\bar K}(u_1,u_2)\big)
&={\rm cs}(u_1-u_2,k')
=k\,{\rm sc}\big({\rm K}(k')+u_2-u_1,k'\big)
\end{eqnarray}
where
\begin{equation}
k'=\sqrt{1-k^2},\qquad{\rm sc}(v,k)={\rm sn}(v,k)/{\rm cn}(v,k)=
1/{\rm cs}(v,k)
\end{equation}
and ${\rm K}(k)$ is the complete elliptic integral of the first kind.

We note that $K$ and ${\bar K}$ are interchanged if we replace $u_1$ by
$u_2\pm{\rm K}(k')$ and $u_2$ by $u_1$. In other words, flipping the
orientation of a rapidity line $j$ is equivalent to changing its rapidity
variable $u_j$ to $u_j\pm{\rm K}(k')$.

\subsection{Two-point correlation functions}

$Z$-invariance implies \cite{BaxZI} that the pair correlation only depends
on elliptic modulus $k$ and the values of the $2m$ rapidity variables
$u_1,\ldots, u_{2m}$ that pass between the two spins, implying the
existence of an infinite set of universal functions
$g_2,g_4,\ldots, g_{2m},\ldots$ such that for any permutation P and
rapidity shift $v$
\begin{equation}
\langle\sigma\sigma'\rangle=g_{2m}(k;{\bar u}_1,\ldots,{\bar u}_{2m})
=g_{2m}(k;{\bar u}_{{\rm P}(1)}+v,\ldots,{\bar u}_{{\rm P}(2m)}+v).
\end{equation}
Here ${\bar u}_j=u_j$ if the $j$th rapidity line passes between the two
spins $\sigma$ and $\sigma'$ in a given direction and
${\bar u}_j=u_j+{\rm K}(k')$ if it passes in the opposite direction
\cite{AP-ZI}.

If two of the rapidity variables passing between the two spins differ by
${\rm K}(k')$, they can be viewed as belonging to a single rapidity line
moving back and forth between these two spins, {\it i.e.} \cite{BaxZI},
\begin{equation}
g_{2m+2}\big(k;{\bar u}_1,\ldots,{\bar u}_{2m},
{\bar u}_{2m+1},{\bar u}_{2m+1}+{\rm K}(k')\big)
=g_{2m}(k;{\bar u}_1,\ldots,{\bar u}_{2m}).
\end{equation}

\subsection{Quadratic identity for pair correlation}

\begin{figure}[tbh]
\begin{center}
\includegraphics[width=0.6\hsize]{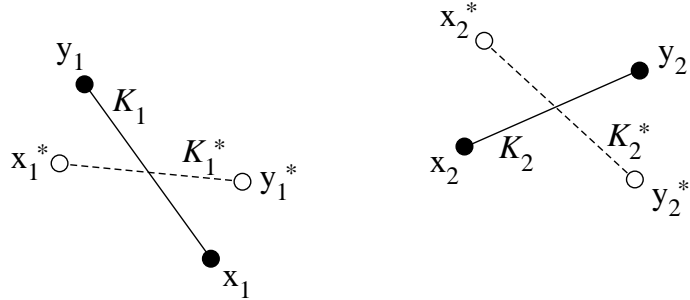}
\end{center}

\caption{At crossing 1 of two rapidity lines we have spin sites
${\rm x}_1$ and ${\rm y}_1$ together with dual spin sites
${\rm x}^{\ast}_1$ and ${\rm y}^{\ast}_1$. At crossing 2 we have
positions ${\rm x}_2$ and ${\rm y}_2$ together with
${\rm x}^{\ast}_1$ and ${\rm y}^{\ast}_1$ oriented similarly.}
\label{fig2}
\end{figure}

For general planar Ising models we can derive a quadratic relation
between pair correlation functions \cite{P1980}. For the situation
in Fig.~\ref{fig2} we have
\begin{eqnarray}
&\sinh(2K_1)\sinh(2K_2)
\big\{ 
\langle\sigma_{{\rm x}_1}\sigma_{{\rm x}_2}\rangle
\langle\sigma_{{\rm y}_1}\sigma_{{\rm y}_2}\rangle-
\langle\sigma_{{\rm x}_1}\sigma_{{\rm y}_2}\rangle
\langle\sigma_{{\rm y}_1}\sigma_{{\rm x}_2}\rangle
\big\} \cr\cr
&\qquad\qquad\qquad+\big\{
\langle\sigma_{{\rm x}_1^\ast}\sigma_{{\rm x}_2^\ast}\rangle^\ast
\langle\sigma_{{\rm y}_1^\ast}\sigma_{{\rm y}_2^\ast}\rangle^\ast-
\langle\sigma_{{\rm x}_1^\ast}\sigma_{{\rm y}_2^\ast}\rangle^\ast
\langle\sigma_{{\rm y}_1^\ast}\sigma_{{\rm x}_2^\ast}\rangle^\ast
\big\}=0,
\end{eqnarray}
with two arbitrary nearest-neighbour pairs of spins at the sites
$\{{\rm x}_1,{\rm y}_1\}\ne\{{\rm x}_2,{\rm y}_2\}$, and corresponding
nearest-neighbour pairs of dual spins at 
$\{{\rm x}_1^\ast,{\rm y}_1^\ast\}$ and 
$\{{\rm x}_2^\ast,{\rm y}_2^\ast\}$, whereas
$\sinh(2K_i)\sinh(2K_i^\ast)\equiv1$, ($i=1,2$). Orientations have to be
consistent as in the picture, otherwise the plus changes to a minus.

Restricted to $Z$-invariant Ising models, the quadratic identity reduces
to
\begin{eqnarray}
&k^2{\rm sc}(u_2-u_1,k'){\rm sc}(u_4-u_3,k') \cr
&\quad\times\big\{g(u_1,u_2,u_3,u_4,\cdots)\,g(\cdots)
-g(u_1,u_2,\cdots)\,g(u_3,u_4,\cdots)\big\} \cr\cr
&+\big\{g^\ast(u_1,u_3,\cdots)\,g^\ast(u_2,u_4,\cdots)
-g^\ast(u_1,u_4,\cdots)\,g^\ast(u_2,u_3,\cdots)\big\}=0,
\end{eqnarray}
with ``$\cdots$" short-hand for all other rapidity variables
$u_5,u_6,\cdots$, common to all $g$'s and $g^\ast$'s
(passing between all eight sites).

Knowing $g(u,u,\cdots,u)$ and $g(v,u,\cdots,u)$, with all or all-but-one
of the rapidities equal, all other $g$'s can be calculated by recurrence.
Therefore, the knowledge of the diagonal and next-to-diagonal pair
correlations in the uniform asymmetric ($K\ne{\bar K}$)
square-lattice Ising model suffices \cite{APMO1,APMO2}.

Jin has found the scaling form for the general $Z$-invariant case in the
critical regime $k\approx1$ \cite{AJP,APMO2}. From this we can find the
first two terms of the susceptibility in several lattices confirming
Guttmann's extended lattice-lattice scaling \cite{APplanar,APquasi}.

\subsection{Summary of findings}

We can make the couplings $J$ and/or the lattice aperiodic. We find the
following \cite{AJP,APMO1,APferro,APpenta}:

\begin{itemize}
\item{Periodic lattice with periodic couplings: Periodic
$\chi({\bf q})$, with peaks at sites commensurate with reciprocal lattice,
becoming sharper and sharper as $T\to T_{\rm c}$. This includes
fully-frustrated cases.}
\item{Periodic lattice with ferromagnetic couplings varying
quasiperiodically: Periodic $\chi({\bf q})$, with peaks at reciprocal
lattice sites, sharper and sharper as $T\to T_{\rm c}$}
\item{Periodic lattice with mixed FM and AFM couplings
quasiperiodically arranged: Periodic $\chi({\bf q})$, with more and more
incommensurate peaks within unit cell as $T\to T_{\rm c}$}
\item{Quasiperiodic lattice: Quasiperiodic $\chi({\bf q})$, with
more and more peaks visible closer to $T_{\rm c}$}
\end{itemize}

For $Z$-invariant lattices, we can evaluate $\chi({\bf q})$ numerically to
high accuracy from the recurrence relations for the pair correlations.
However, the structure is clearer in density plots \cite{AJP,APMO1}.

\subsection{Generalized Fibonacci Ising lattices}

We can assign the couplings according to quasiperiodic sequences in
horizontal,\break vertical, and/or diagonal directions.
We can use \cite{APferro} de Bruijn's generalized Fibonacci sequences,
assigning different couplings according to the sequence of zeros and ones
\begin{equation}
p_j(n)\equiv\lfloor \gamma+(n+1)/\alpha_j\rfloor-\lfloor
\gamma+n/\alpha_j\rfloor
\end{equation}
with
\begin{equation}
\alpha_j\equiv{1\over2}\big[(j+1)+\sqrt{(j+1)^2+4}\,\big]
\end{equation}
We found that such ferromagnetic cases differ very little from periodic
cases.

For the mixed ferro/antiferro case, the $\chi({\bf q})$ depends strongly
on the sequence chosen, even for the simplest examples just adding signs
to the couplings of the Onsager square lattice model by gauge transform
\cite{APferro}.

\subsection{Pentagrid Ising lattices}

Following Korepin, we can use de Bruijn's pentagrid for the rapidity lines
leading to a spin model and its dual taking alternating sites of a Penrose
tiling \cite{APpenta}. By the quadratic recurrence relations we can compute
a big collection of pair correlations. Then, in order to calculate
$\chi({\bf q})$, we have developed a novel way of determining the pair
probability of local environments on a Penrose tiling, which can also be
used once more detailed results for pair correlations in e.g.\ the
eight-vertex model or the chiral Potts model become available. Full
details will be published elsewhere \cite{APpenta}.

\section{Fully-frustrated square-lattice Ising model}
\begin{figure}[tbh]
\begin{center}
\includegraphics[width=0.6\hsize]{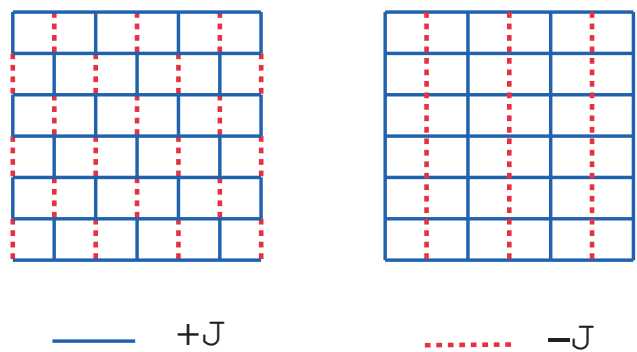}
\end{center}

\caption{Two versions of the fully-frustrated square-lattice Ising
model with ferromagnetic couplings $+J$ and antiferromagnetic couplings
$-J$: (a) Checkerboard version on the left; (b) version periodic in
vertical direction on the right.}
\label{fig3}
\end{figure}

There are two common ways to define a fully-frustrated Ising model
on a square lattice. One way is the checkerboard version of coloring
the faces of the lattice alternatingly black and white and taking
three of the couplings around each black square $+J$ and the fourth one
$-J$. The other way is to take all horizontal couplings $+J$ and the
couplings in vertical columns alternatingly $+J$ and $-J$. This is
illustrated in Fig.~\ref{fig3}.

\begin{figure}[tbh]
\begin{center}
\includegraphics[width=0.3\hsize]{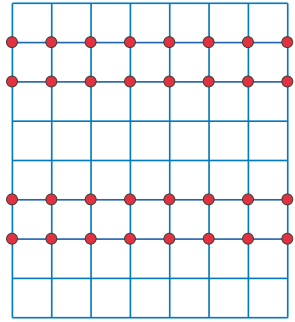}
\end{center}
\vskip6pt

\caption{Gauge equivalence of the checkerboard and vertically periodic
cases: Flip the signs of the  red spins to go from the one to the other,
or back.}
\label{fig4}
\end{figure}

It must be noted that these two versions do not define fundamentally
different models. They are, in fact, related with each other by a simple
gauge transformation of flipping the signs of each second horizontal pair
of rows of spins, as is indicated in Fig.~\ref{fig4}.
Of course, there are many other ways of flipping subsets of the spins and
one can define infinitely many other related less regular frustrated
models.

There are several approaches to the explicit evaluation of pair
correlations and susceptibilities in this model.

\subsection{Sum out every other spin}
\begin{figure}[tbh]
\begin{center}
\includegraphics[width=0.2\hsize]{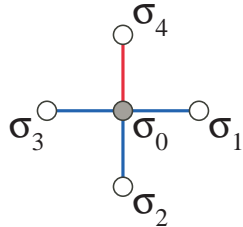}
\end{center}

\caption{Spin $\sigma_0$ to be ``decimated out" with its four neighbours.}
\label{fig5}
\end{figure}

One approach is to first sum out every other spin in order to arrive
at an effective Baxter eight-vertex model \cite{F,FF}, with diagonal
couplings $\hat J$ and $\hat J'$ and four-spin coupling $\hat J_4$.
If one does this for the checkerboard case, see also Fig.~\ref{fig5}, one
finds
\begin{equation}
{\rm e}^{4\hat K}={1\over\sqrt{2S^2+1}},\quad
{\rm e}^{4\hat K'}=\sqrt{2S^2+1},\quad
{\rm e}^{4\hat K_4}={S^2+1\over\sqrt{2S^2+1}}
\end{equation}
with $\hat K\equiv\hat J/k_{\rm B}\hat T$,
$\hat K'\equiv\hat J'/k_{\rm B}\hat T$ and
$\hat K_4\equiv\hat J_4/k_{\rm B}\hat T$, while using the short-hand
notation
\begin{equation}
S\equiv\sinh{2J\over k_{\rm B}T}\equiv\sinh2K
\end{equation}
In terms of Baxter's $a$, $b$, $c$, $d$, this means
\begin{equation}
a=b=\sqrt{S^2+1},\quad c=1,\quad d=\sqrt{2S^2+1}
\end{equation}
Note that only diagonal and four-spin interactions are left. All other
interactions cancel out.

This determines the mapping of Boltzmann weights. For correlation
functions we have also to find the mapping of the spins involved. Spins
$\sigma_1$, $\sigma_2$, $\sigma_3$, $\sigma_4$ in Fig.~\ref{fig5} are left
unchanged. However, the central spin $\sigma_0$ in Fig.~\ref{fig5} has to
be replaced after ``decimation summation" according to
\begin{equation}
\sigma_0\quad\longrightarrow\quad
{S\,(\sigma_1\!+\!\sigma_3\!+\!\sigma_2\!-\!\sigma_4)\over2\sqrt{S^2+1}}
\Big(1-{S^2(1-\sigma_1\sigma_3\sigma_2\sigma_4)\over2(2S^2+1)}\Big)
\end{equation}

\subsection{Partial duality transform}
After this, we first flip the sign of each $\sigma_2\sigma_4$,
interchanging
$a\leftrightarrow d$, $b\leftrightarrow c$. We then have
\begin{equation}
{\rm e}^{4\tilde K}={\rm e}^{4\tilde K'}=\sqrt{2S^2+1},\qquad
{\rm e}^{4\tilde K_4}={\sqrt{2S^2+1}\over S^2+1},\qquad
S\equiv\sinh{2J\over k_{\rm B}T}
\end{equation}

Next, we apply a partial Kramers--Wannier duality transform on every other
spin \cite{F,FF}. This leads to an Ashkin--Teller model, with new spins
$\tau$ on the same positions as a quarter of the original spins
$\sigma$ and three quarters of the original positions empty.
All original spins $\sigma$ are now expressed in terms of $\sigma$'s and
$\tau^*$'s  (disorder variables, equal to other original $\sigma$'s).

Finally, replacing
$\sigma\longrightarrow\sigma\tau$, $\tau\longrightarrow\tau$,
$\sigma^*\longrightarrow\sigma^*$, $\tau^*\longrightarrow\sigma^*\tau^*$,
the Ashkin--Teller model factorizes into two Ising models, at dual
temperatures
$K_{\sigma}=\beta_{\sigma}J_0$, $K_{\tau}=\beta^*_{\sigma}J_0$, with
\begin{equation}
\sinh(2K_{\sigma})={S^2\over S^2+1+\sqrt{2S^2+1}}
={1\over\sinh(2K_{\tau})}
\equiv\sqrt{k}
\end{equation}

\subsection{Difference equations for the correlation functions}
Writing $C(m,n)\equiv
\langle\sigma_{0,0}\sigma_{m,n}\rangle_{K_\sigma}^{\phantom{*}}$ and
$\bar C(m,n)\equiv
\langle\tau_{0,0}\tau_{m,n}\rangle_{K_\tau}^{\phantom{*}}$ for the two
resulting independent Ising models, we can determine these recursively
\cite{P1980,APMO2} from
\begin{eqnarray}
&\big[C(m,n\!+\!1)C(m,n\!-\!1)-C(m,n)^2\big]\nonumber\\
&\quad+k\,\big[\bar C(m\!+\!1,n)\bar C(m\!-\!1,n)-\bar C(m,n)^2\big]=0\\
&\big[C(m\!+\!1,n)C(m\!-\!1,n)-C(m,n)^2\big]\nonumber\\
&\quad+k\,\big[\bar C(m,n\!+\!1)\bar C(m,n\!-\!1)-\bar C(m,n)^2\big]=0\\
&\big[C(m,n)C(m\!+\!1,n\!+\!1)-C(m\!+\!1,n)C(m,n\!+\!1)\big]=\nonumber\\
&\qquad k\,\big[\bar C(m,n)\bar C(m\!+\!1,n\!+\!1)-
\bar C(m\!+\!1,n)C(m,n\!+\!1)\big]\\
&\sqrt{k}\,\big[C(m\!+\!1,n)\bar
C(m\!-\!1,n)+C(m\!-\!1,n)\bar C(m\!+\!1,n)\nonumber\\
&\quad+C(m,n\!+\!1)\bar C(m,n\!-\!1)
+C(m,n\!-\!1)\bar C(m,n\!+\!1)\big]\nonumber\\
&\quad=\,(k\!+\!1)\,C(m,n)\bar C(m,n)
\end{eqnarray}
with
\begin{eqnarray}
&C(0,0)=\bar C(0,0)=1,\quad C(1,0)+\sqrt{k}\,\bar C(0,1)=\sqrt{k\!+\!1}\\
&C(m,n)=C(n,m)=C(|m|,|n|)\nonumber\\
&\bar C(m,n)=\bar C(n,m)=\bar C(|m|,|n|)
\end{eqnarray}

\subsection{Pair correlation function of fully-frustrated model}
Finally, the pair correlation functions of the original model of
Fig.~\ref{fig3}$\,$(a) are given by
\begin{eqnarray}
&\langle\sigma^{\vphantom{\dagger}}_{kl}\,
\sigma^{\vphantom{\dagger}}_{k+2m,l+2n}\rangle\
=(-1)^n\,C(m,n)\,\bar{C}(m,n)\\
&\langle\sigma^{\vphantom{\dagger}}_{kl}\,
\sigma^{\vphantom{\dagger}}_{k+2m-1,l+2n-1}\rangle
=0\\
&\langle\sigma^{\vphantom{\dagger}}_{kl}\,
\sigma^{\vphantom{\dagger}}_{k+2m-1,l+2n}\rangle\nonumber\\
&\qquad={(-1)^n\,S\over2\sqrt{2S^2+1}}\,
\big(C(m-1,n)\,\bar{C}(m,n)+C(m,n)\,\bar{C}(m-1,n)\big)\\
&\langle\sigma^{\vphantom{\dagger}}_{kl}\,
\sigma^{\vphantom{\dagger}}_{k+2m,l+2n-1}\rangle\nonumber\\
&\qquad={\pm(-1)^n\,S\over2\sqrt{2S^2+1}}\,
\big(C(m,n-1)\,\bar{C}(m,n)+C(m,n)\,\bar{C}(m,n-1)\big)
\end{eqnarray}
In the last equation we have to choose plus
($+$) if $\min(k+l,k+2m+l+2n-1)$ = even
and minus
($-$) if $\min(k+l,k+2m+l+2n-1)$ = odd.
Therefore, its contribution to $\chi(q_x,q_y)$ averages out to zero.
If we had started from the vertically periodic case Fig.~\ref{fig3}$\,$(b),
we must omit all three $(-1)^n$ factors in the above.

Obviously both models have a periodic $\chi(q_x,q_y)$ with commensurate
peaks only, as can be illustrated with density plots. Using the results
of this subsection it is also not difficult to derive a long series
expansion along the lines of the Ising model work \cite{ONGP1}.

\subsection{Other approaches}

The square-lattice fully-frustrated Ising model is dual to the Ising model
in field ${\rm i}\pi k_{\rm B}T/2$ introduced by Lee and Yang. For this
case it in convenient to view a square of four spins as a vertex of a
16-vertex model and then to apply gauge transforms \cite{AP-ZI}. Now the
model decouples as a product of two Ising models at the same temperature.
For the special case of the square-lattice dimer model these two models are
at the critical temperature and the explicit formula for the two-point
function has been published \cite{APff} and the monomer-monomer
correlation function and its Fourier transform $\chi(q_x,q_y)$ have been
studied in some detail by Kong \cite{Kong}.

\subsection{Final remarks: Checkerboard Ising and chiral Potts}
The physical free-fermion model corresponds to a checkerboard Ising model
\cite{BaxFF} with real or imaginary elliptic modulus $k$. In general, the
correlation functions satisfy a 2-by-2 matrix generalization \cite{Jxy} of
the difference equations used above, allowing incommensurate solutions.

For the two-dimensional integrable $N$-state chiral Potts model a similar
mapping to an $N$-state generalization of the free-fermion 8-vertex
model exists \cite{BPA}. This model is a most natural generalization of
Onsager's two-dimensional Ising model to more than two states per spin.
The pair-correlation functions have not yet been solved for this model;
only the conjecture for its order parameters (one-point functions) has been
proved recently by Baxter \cite{Baxter}. The model is on a submanifold in
the commensurate phase of the more general chiral Potts model. The full
chiral Potts model also has incommensurate phases. However, Jin's results
\cite{Jin} do not seem to support the presence of a Lifshitz point in the
classical two-dimensional model.

\section{Overlapping unit cells}
Gummelt \cite{Gummelt} motivated by physical considerations, has proposed
a description of quasicrystals in terms of overlapping unit cells, with
the regular Penrose tiling described by overlappings of decorated
decagons. We have used a multigrid method based on de Bruijn's work to
produce a new example of 3-dimensional overlapping unit cells,
quasiperiodic in two directions and periodic in the third. Full details
are presented elsewhere \cite{APoverlap}. 

As our construction is based on a multigrid of five grids of parallel
planes, resulting from a projection of the five-dimensional hypercubic
lattice into three dimensions, one may construct aperiodic integrable
three-dimensional models with spectral variables living on the grid
planes and with Boltzmann weights satisfying the tetrahedron equations,
generalizing the pentagrid Ising model construction.

\ack
One of us (JHHP) thanks the organizing committee of the Dunk Island
conference for their kind  invitation. 


\section*{References}
\begin{thebibliography}{99}

%
\bibitem{WMTB76}{
Wu T T, McCoy B M, Tracy C A and Barouch E 1976
Spin-spin correlation functions for the two-dimensional
Ising model: Exact theory in the scaling region
\PR B {\bf 13} 316--374}
%
\bibitem{ONGP1}{
Orrick W P, Nickel B G, Guttmann A J and Perk J H H 2001
Critical behavior of the two-dimensional Ising susceptibility
\PRL {\bf 86} 4120--3}
%
\bibitem{ONGP2}{
Orrick W P, Nickel B, Guttmann A J and Perk J H H 2001
The susceptibility of the square lattice Ising model: New developments
{\it J. Stat. Phys.} {\bf 102} 795--841}
%
\bibitem{P1980}{
Perk J H H 1980
Quadratic identities for Ising model correlations
{\it Phys. Lett.} A {\bf 79} 3--5}
%
\bibitem{AJP}{
Au-Yang H, Jin B-Q and Perk J H H 2001
Wavevector-dependent susceptibility in quasiperiodic Ising models
{\it J. Stat. Phys.} {\bf 102} 501--43}
%
\bibitem{APMO1}{
Au-Yang H and Perk J H H 2002
Wavevector-dependent susceptibility in aperiodic planar Ising models
{\it MathPhys Odyssey 2001: Integrable Models and Beyond}
ed Kashiwara M and Miwa T (Birkh\"auser, Boston) pp~1--21}
%
\bibitem{APMO2}{
Au-Yang H and Perk J H H 2002
Correlation functions and susceptibility in the $Z$-invariant Ising model
{\it MathPhys Odyssey 2001: Integrable Models and Beyond}
ed Kashiwara M and Miwa T (Birkh\"auser, Boston) pp~23--48}
%
\bibitem{CHPV}{
Caselle M, Hasenbusch M, Pelissetto A and Vicari E 2002
Irrelevant operators in the two-dimensional Ising model
\JPA {\bf 35} 4861--88}
%
\bibitem{BaxZI}{
Baxter R J 1978
Solvable eight vertex model on an arbitrary planar lattice
{\it Phil. Trans. R. Soc. Lond.} A {\bf 289} 315--46}
%
\bibitem{BaxFF}{
Baxter R J 1986
Free-fermion, checkerboard and $Z$-invariant lattice
models in statistical mechanics
{\it Proc. R. Soc. Lond.} A {\bf 404} 1--33}
%
\bibitem{AP-ZI}{
Au-Yang J and Perk J H H 1987
Critical correlations in a $Z$-invariant inhomogeneous Ising model
{\it Physica} A {\bf 144} 44--104}
%
\bibitem{APplanar}{
Au-Yang H and Perk J H H 2002
New results for susceptibilities in planar Ising models
{\it Int. J. Mod. Phys.} B {\bf 16} 2089--95}
%
\bibitem{APquasi}{
Au-Yang H and Perk J H H 2003
Susceptibility calculations in periodic and
quasiperiodic planar Ising models
{\it Physica} A {\bf 321} 81--9}
%
\bibitem{APferro}{
Au-Yang H and Perk J H H 2005
$Q$-dependent susceptibilities in ferromagnetic
quasiperiodic $Z$-invariant Ising models
{\it Preprint}}
%
\bibitem{APpenta}{
Au-Yang H and Perk J H H 2005
$Q$-dependent susceptibilities in $Z$-invariant pentagrid Ising models
{\it Preprint} cond-mat/0409557}
%
\bibitem{F}{Forg\'acs G 1980
Ground-state correlations and universality in
two-dimensional fully frustrated Ising model
\PR B {\bf 22} 4473--80}

%
\bibitem{FF}{Forg\'acs G and Fradkin E 1981
Anisotropy and marginality in the two-dimensional fully
frustrated Ising model
\PR B {\bf 23} 3442-7}
%
\bibitem{APff}{
Au-Yang H and Perk J H H 1984
Ising correlations at the critical temperature
{\it Phys. Lett.} A {\bf 104A} 131--4}
%
\bibitem{Kong}{
Kong X-P 1987
Wave-Vector Dependent Susceptibility of the
Two-Dimensional Ising Model
{\it Ph D Thesis} (State University of New York at Stony Brook)}
%
\bibitem{Jxy}{
Perk J H H 1980
Equations of motion for the transverse correlations of
the one-dimensional XY-model at finite temperature
{\it Phys. Lett.} A {\bf 79} 1--2}
%
\bibitem{BPA}{
Baxter R J, Perk J H H and Au-Yang H 1988
New solutions of the star-triangle relations for the chiral Potts model
{\it Phys. Lett.} A {\bf 128} 138--42}
%
\bibitem{Baxter}{
Baxter R J 2005
The order parameter of the chiral Potts model
{\it J. Stat. Phys.} {\bf 120} 1--36}
%
\bibitem{Jin}{
Jin B-Q 2001
Some Aspects of the Chiral Potts Model and the Ising Model
{\it Ph D Thesis} (Oklahoma State University, Stillwater)}
%
\bibitem{Gummelt}{
Gummelt P 1996
Penrose tilings as coverings of congruent decagons
{\it Geometriae Dedicata} {\bf 62} 1--17}
%
\bibitem{APoverlap}{
Au-Yang H and Perk J H H 2005
Overlapping unit cells in 3-d quasicrystal structure
{\it Preprint} cond-mat/0507117}
%

\end{thebibliography}
\end{document}